\documentstyle[aps,multicol,epsfig]{revtex}

\newcommand{\<}{\langle}
\renewcommand{\>}{\rangle}

\newcommand{\be}{\begin{equation}}
\newcommand{\ee}{\end{equation}}
\newcommand{\bea}{\begin{eqnarray}}
\newcommand{\eea}{\end{eqnarray}}

\newcommand{\sign}{\rm{sign}}
\renewcommand{\deg}{^\circ}

\begin{document}

%%%%%%%%%%%%%%%%%%%%%%%%%%%%%%%%%%%%%%%%%%%%%%%%%%%%%%%%%%%%%%%%%%%%%%%%%%%%%%%
% Title

\title{Universal quantum computation with two-level trapped ions} 
\author{Andrew M. Childs and Isaac L. Chuang}
\address{IBM Almaden Research Center, 650 Harry Road, San Jose, CA 95120}
\date{14 August 2000}
\maketitle

%%%%%%%%%%%%%%%%%%%%%%%%%%%%%%%%%%%%%%%%%%%%%%%%%%%%%%%%%%%%%%%%%%%%%%%%%%%%%%%
% Abstract

\begin{abstract}
Although the initial proposal for ion trap quantum computation made use of
an auxiliary internal level to perform logic between ions, this resource is
not necessary in principle.  Instead, one may perform such operations
directly using sideband laser pulses, operating with an arbitrary
(sufficiently small) Lamb-Dicke parameter.  We explore the potential of this
technique, showing how to perform logical operations between the internal
state of an ion and the collective motional state and giving explicit
constructions for a controlled-{\sc not} gate between ions.
\end{abstract}


% PACS 99
% 03.67.Lx Quantum computation
% 32.80.Lg Mechanical effects of light on atoms, molecules, and ions

\begin{multicols}{2}
\narrowtext

%%%%%%%%%%%%%%%%%%%%%%%%%%%%%%%%%%%%%%%%%%%%%%%%%%%%%%%%%%%%%%%%%%%%%%%%%%%%%%%
\section{Introduction}

Ion trap quantum computation, first introduced by Cirac and
Zoller~\cite{CZ95}, is a potentially powerful technique for the storage and
manipulation of quantum information~\cite{CZ95,Win97,Ste97,Hug98}.  In this
scheme, information is stored in the spin states of an array of trapped ions
and manipulated using laser pulses.  Reasonably long coherence times can be
achieved, compared to achievable switching rates~\cite{Div95}, and
individual qubits can be addressed through spatial separation of the ions.
Experimental implementations of this scheme have succeeded in performing
simple two-qubit logic gates~\cite{Mon95,Roos99} and preparing entangled
states~\cite{Tur98,Sac00}.

An ion trap quantum computer may be modeled as a collection of $N$ particles
with spin $1 \over 2$ in a one-dimensional harmonic potential.  Laser pulses
incident on the ions can be tuned to simultaneously cause internal spin
transitions and vibrational (phonon) excitations, thus allowing local
internal states to be mapped into shared phonon states.  In this manner,
quantum information can be communicated between any pair of ions and logic
gates can be performed.

In this paper, we consider an interesting question which arises in this
scenario: what is the simplest internal spin state structure required by
each ion?  In the original Cirac-Zoller formulation, ions with three levels
are required.  However, as the Cirac-Zoller method is generalized to other
physical systems, such as neutral trapped atoms\cite{Pel95} or quantum dots
in an electromagnetic cavity\cite{Ima99}, it has become highly desirable to
determine whether just {\em two} internal levels are sufficient for
performing universal quantum computation.

Previously, Monroe {\em et al.}\ have shown how a controlled-{\sc not} can
be performed between an ion and a phonon state using only two-level
ions~\cite{Mon97}.  Their method depends on fine-tuning of the Lamb-Dicke
parameter, which relates the laser frequncy to the scale of the ions' wave
functions, to cancel unwanted side effects.  Furthermore, they do not allow
the Lamb-Dicke parameter to be arbitrarily small; the lowest value quoted
in~\cite{Mon97} is 0.316.

Here, we provide a general and accessible technique for performing universal
logic between ions with only two internal levels.  This scheme operates with
any sufficiently small value of the Lamb-Dicke parameter, and also
introduces new ways to utilize (or avoid) particular phonon states while
performing quantum logic gates.  We begin in Section~\ref{sec:theory} by
presenting the allowed operations using the usual Jaynes-Cummings model for
spin-boson interactions.  We then describe a multiple-pulse construction for
the controlled-{\sc not} gate in Section~\ref{sec:cnot} and give further
constructions in Section~\ref{sec:other} before concluding with some
possible extensions of the work.

%%%%%%%%%%%%%%%%%%%%%%%%%%%%%%%%%%%%%%%%%%%%%%%%%%%%%%%%%%%%%%%%%%%%%%%%%%%
\section{Theoretical ion trap model}
\label{sec:theory}

\begin{figure}
\begin{center}
\psfig{figure=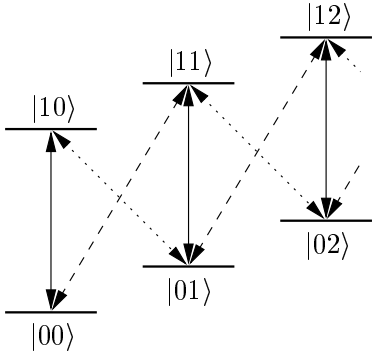,width=2in}
\end{center}
\caption{Energy level diagram for a single ion's nuclear spin state along
with the motional mode, showing only the lowest three motional states.  We
denote the state of a particular ion using the ket $|p q\>$, where $p \in
\{0,1\}$ is the nuclear spin state and $q \in \{0,1,2,\ldots\}$ is the
motional state.  Transitions at the on-resonance frequency $\omega_0$ are
shown with solid lines.  The first blue ($\omega_0+\omega_z$) and red
($\omega_0-\omega_z$) sideband transions are shown using dashed and dotted
lines, respectively.  Higher-order transitions are suppressed for clarity.}
\label{fig:energy}
\end{figure}

The energy level diagram of one ion, including the motional state, is shown
in Fig.~\ref{fig:energy}.  For the sake of definiteness, we may think of the
motional levels as corresponding to the center of mass degree of freedom.
However, they could just as well correspond to another mode of oscillation,
such as the ``breathing'' mode of a pair of ions.  In practice, one might
wish to choose a mode other than the center of mass to achieve reduced
susceptibility to decoherence~\cite{King98}.

The Hamiltonian of this system is $H_0=\hbar \omega_0 {\sigma_z \over 2} +
\hbar \omega_z a^\dagger a$, where $\sigma_z$ is a Pauli spin operator for
the nuclear spin and $a$ annihilates a phonon.  Throughout this paper, we
work in the frame of this Hamiltonian.  Turning on the electromagnetic field
of a laser gives an interaction Hamiltonian
\be
\label{eq:interaction}
H_I = - \vec\mu \cdot \vec B
\,,
\ee
where $\vec \mu = \mu \vec \sigma/2$ is the magnetic moment of the ion and
$\vec B=B \hat x \cos(kz-\omega t+\Phi)$ is the magnetic field produced by
the laser.  Here $z=z_0(a+a^\dagger)$, where $z_0=\sqrt{\hbar/2Nm\omega_z}$
is a characteristic length scale for the motional wave functions and $m$ is
the mass of an ion.

We consider the regime in which $\eta \equiv k z_0 \ll 1$.  In this regime,
we may determine the effect of a laser pulse at a specific frequency
$\omega$ by expanding Eq.~(\ref{eq:interaction}) in powers of $\eta$ and
neglecting rapidly rotating terms.  Pulsing on resonance ($\omega =
\omega_0$) allows one to perform the transformation
\be
R(\theta,\phi) 
= \exp\left[i{\theta \over 2} ( e^{i\phi} \sigma^+ + e^{-i\phi} \sigma^- )
      \right]
\,,
\ee
allowing one to do arbitrary single-qubit operations on an ion's internal
state.  Pulsing at the $n$th blue sideband frequency
($\omega=\omega_0+n\omega_z$) gives~\cite{SigmaNote}
\be
\label{eq:blue}
R^+_n(\theta,\phi) 
= \exp\left[i {\theta \over 2} ( e^{i\phi}  \sigma^+ a^n
                               + e^{-i\phi} \sigma^- a^{\dagger n} ) \right]
\ee
and pulsing at the $n$th red sideband frequency ($\omega=\omega_0-n\omega_z$)
gives 
\be
\label{eq:red}
R^-_n(\theta,\phi) 
= \exp\left[i {\theta \over 2} ( e^{i\phi}  \sigma^+ a^{\dagger n}
                               + e^{-i\phi} \sigma^- a^n ) \right]
\,.
\ee
Here, $\sigma^\pm = (\sigma_x \pm i \sigma_y)/2$ act on the internal state
of the ion.  In each case, the parameter $\theta$ depends on the strength
and duration of the pulse and $\phi$ depends on its phase.  For an $n$th
order transition of duration $t$, $\theta$ is given by
\be
\theta = - { \mu B t \eta^n \over 2 \hbar n!}
\,.
\ee
Also,
\be
\phi=\Phi+(n \,{\rm mod}\, 4){\pi \over 2}
\,.
\ee

%%%%%%%%%%%%%%%%%%%%%%%%%%%%%%%%%%%%%%%%%%%%%%%%%%%%%%%%%%%%%%%%%%%%%%%%%%%
\section{CNOT using the first motional sideband}
\label{sec:cnot}

We suppose that the state of the ion trap quantum computer begins in the
motional $|0\>$ state.  We may excite higher-order motional levels using
$R^\pm_n$ so long as we return to the motional ground state at the end of
the gate.  It is simplest to consider cases where only a few higher-order
levels are relevant.  For a particular ion, we refer to the subspace spanned
by $\{|00\>,|01\>,|10\>,|11\>\}$ (using the notation of
Figure~\ref{fig:energy}) as the {\em computational subspace} (CS).

First, we note that for the special values $\theta = j \pi \sqrt2$, where
$j$ is an integer, $R^\pm_1$ preserve the CS --- that is, they map states
within the CS to other states within the CS.  For example,
\be
\label{eq:blueCS}
R^+_1(\pi \sqrt2,-\pi/2)=\left(
\matrix{\cos{\pi \over \sqrt2} & 0 & 0 & \sin{\pi \over \sqrt2} \cr
        0 & -1 & 0 & 0 \cr
        0 &  0 & 1 & 0 \cr
        -\sin{\pi \over \sqrt2} & 0 & 0 & 
         \cos{\pi \over \sqrt2}} \right)
\,.
\ee

Next, note that it is easy to produce another gate which preserves the CS by
conjugation.  For example, consider the gate $G=G_{\rm in} \otimes G_{\rm
out}$, where $G_{\rm in}$ has support only on $\{|00\>,|10\>,|11\>\}$ and
$G_{\rm out}$ acts on the rest of the space.  Since $R^+_1$ does not connect
these subspaces, conjugating this gate by $R^+_1$ produces another gate
which preserves the CS.  In other words,
\be
R^+_1(-\theta,\phi) G R^+_1(\theta,\phi)
\ee
preserves the CS.  Using this observation, we can easily diagonalize
Eq.~(\ref{eq:blueCS}), giving the gate
\bea
P &=& R^+_1(-\pi/2,0) R^+_1(\pi \sqrt2,-\pi/2) R^+_1(\pi/2,0)
\nonumber \\
\label{eq:diaggate}
&=& {\rm diag}(e^{i\pi/\sqrt2},-1,1,e^{-i\pi/\sqrt2})
\,.
\eea
This gate uses first-order sideband pulses for a total duration of
$(1+\sqrt{2})\pi$, only slightly longer than the $2\pi$ duration required
for the equivalent step in the Cirac-Zoller scheme.

With this diagonal gate and $R^-_1$, it is straightforward to construct a
controlled-{\sc not} gate between two ions using a sequence similar to the
original Cirac-Zoller construction.  Note that $R^-_1(\pi,-\pi/2)$ allows us
to interchange $|01\>$ and $|10\>$ (up to a phase), performing
\be
\left(\matrix{ 1 &  0 & 0 \cr 
               0 &  0 & 1 \cr 
               0 & -1 & 0 }\right)
\ee
on the subspace $\{|00\>,|01\>,|10\>\}$.  Assuming the motional state is
initially $|0\>$, this corresponds to a swap between the state of the ion
and the motional state.  In fact, we can use a version of this gate with
arbitrary phase to do logic between ions.  Thus we can perform a
controlled-{\sc not} from ion $j$ (the control) to ion $k$ (the target)
using
\bea
\label{eq:cnot}
CNOT_{jk} =&& Z_j(-\pi / (2\sqrt2)) R^-_{1j}(\pi,\phi) \\ \nonumber
           && H_k P_k Z_k(-\pi / (2\sqrt2)) H_k R^-_{1j}(\pi,\phi)
\,,
\eea
for any value of $\phi$, where a subscript denotes which ion is acted on and
we have introduced the single-qubit gates
\be
Z(\phi) = \left( \matrix{e^{i \phi} & 0 \cr 0 & e^{-i \phi}} \right)
\,,\quad
H = {1 \over \sqrt2} \left( \matrix{1 & 1 \cr 1 & -1} \right)
\,.
\ee

%%%%%%%%%%%%%%%%%%%%%%%%%%%%%%%%%%%%%%%%%%%%%%%%%%%%%%%%%%%%%%%%%%%%%%%%%%%
\section{Other gate constructions}
\label{sec:other}

Although the construction by which we arrived at Eq.~(\ref{eq:diaggate}) is
relatively straightforward, there are other ways to construct gates which
preserve the CS.  In fact, it is possible to perform a generalization of
Eq.~(\ref{eq:diaggate}) by considering the form
\bea
U(\phi) = && R^+_1(-\alpha,\pi/2) R^+_1(-\beta,\gamma) R^+_1(2\phi,\delta) 
          \\ \nonumber
          && R^+_1(\beta,\gamma)  R^+_1(\alpha,\pi/2)
\,.
\eea
Making the choices
\bea
\cos \alpha &=& \cos \sqrt 2 \alpha \\
\cos \beta  &=& \cos \sqrt 2 \beta \\
\sign(\sin\alpha) \sign(\sin\beta) &=& 
  \sign(\sin \sqrt 2 \alpha) \sign(\sin \sqrt 2 \beta) \\
\delta-\gamma &=& \sin^{-1}\left(\cos\alpha \over \sin \beta\right) \\
\gamma &=& \tan^{-1}[-\cos\beta \tan(\delta-\gamma)]
\eea
results in
\be
\label{eq:generaldiaggate}
U(\phi)={\rm diag}(e^{i\phi},e^{\pm i\sqrt{2}\phi},1,e^{-i\phi})
\,.
\ee
For example, there is such a solution with $\alpha \approx 298.2\deg$,
$\beta \approx 149.1\deg$, $\gamma \approx 63.87\deg$, and $\delta \approx
131.0\deg$.

The source of the factors of $\sqrt2$ in gates derived from the first-order
sideband is the matrix element $\<1|a|2\>=\sqrt2$ for harmonic oscillator
states.  However, note that
\be
\<1|a^3|4\>=2\<0|a^3|3\>=2\sqrt6
\,.
\ee
The integer ratio of these matrix elements suggests that third-order
transitions may be used to create simpler diagonal gates.  Indeed, we find 
\be
R^-_3(2\pi/\sqrt6,\phi')={\rm diag}(1,1,-1,1)
\ee
for any value of $\phi'$.  This differs from a controlled-$Z$ gate by only
single-qubit operations, so we can simply use the Cirac-Zoller construction
to produce a CNOT between ions:
\bea
\label{eq:highercnot}
CNOT_{jk} =&& H_k Z_j(\pi/2) R^-_{1j}(\pi,\phi) R^-_{3k}(2\pi/\sqrt6,\phi') 
           \\ \nonumber
           && Z_k(-\pi/2) R^-_{1j}(\pi,\phi) H_k
\eea
Here, we require fewer pulses than in Eq.~(\ref{eq:cnot}).  However, note
that the third-order sideband pulse must be longer by a factor of order
$\eta^{-2}$ than the first-order sideband pulses for the same laser
intensity.

Most of these results generalize to other choices of the computational
subspace.  For example, consider using $\{|00\>,|02\>,|10\>,|12\>\}$.
Analogous to Eq.~(\ref{eq:cnot}), we find
\bea
CNOT_{jk} =&& Z_j(-\pi/(2\sqrt6)) R^-_{2j}(\pi/\sqrt2,\phi) H_k \\ \nonumber
           && R^+_{2k}(-\pi/(2\sqrt2),0) 
              R^+_{2k}(\pi/\sqrt3,-\pi/2) \\ \nonumber
           && R^+_{2k}(\pi/(2\sqrt2),0) Z_k(-\pi/(2\sqrt6)) \\ \nonumber
           && H_k R^-_{2j}(\pi/\sqrt2,\phi)
\,.
\eea
Similarly, analogous to Eq.~(\ref{eq:highercnot}) (exploiting the integral
relationship $\<2|a^7|9\>=6\<0|a^7|7\>$),
\bea
CNOT_{jk} =&& H_k Z_j(\pi/2) R^-_{2j}(\pi/\sqrt2,\phi)  \\ \nonumber
           && R^-_{7k}(\pi/(6\sqrt{35}),\phi') Z_k(-\pi/2) \\ \nonumber
           && R^-_{2j}(\pi/\sqrt2,\phi) H_k
\,.
\eea

Finally, we wish to point out that it is conceivable to treat the lowest two
motional levels as an additional qubit, rather than simply an intermediary
for logic, if one can perform a true swap operation over the entire
computational subspace.  Note that
\bea
&&R^-_1(2l\pi\sqrt2,-\pi/2) \nonumber \\
&&\quad = \left(\matrix{ 1 & 0 & 0 & 0 \cr
         0 & \cos(l\pi\sqrt2) & \sin(l\pi\sqrt2) & 0 \cr
         0 &-\sin(l\pi\sqrt2) & \cos(l\pi\sqrt2) & 0 \cr
         0 & 0 & 0 & 1 } \right)
\,.
\eea
We may get arbitrarily close to a swap operation by choosing large enough
$l$ with $\cos(l\pi\sqrt2) \approx 0$.  For example, choosing $l=2378$ gives
a swap operation to a precision of about $10^{-3}$.  Although this
observation may not be useful in practice, it shows that universal quantum
logic including a motional qubit is not forbidden in principle.

%%%%%%%%%%%%%%%%%%%%%%%%%%%%%%%%%%%%%%%%%%%%%%%%%%%%%%%%%%%%%%%%%%%%%%%%%%%
\section{Conclusions}

We have demonstrated the possibility of doing ion trap quantum computation
with two-level ions, allowing any sufficiently small value of the Lamb-Dicke
parameter.  While we believe that a construction similar to
Eq.~(\ref{eq:cnot}) will be most useful in practice, we have shown that
there are many other ways to realize a controlled-{\sc not} gate between
ions without using extra levels, some of which might be useful given
appropriate experimental conditions.

Our initial motivation for gates such as Eqs.~(\ref{eq:diaggate}) and
(\ref{eq:generaldiaggate}) came from the theory of composite pulses used in
the art of NMR, in which errors in pulse length or frequency are canceled at
low order using a sequence of pulses.  Thus one may accomplish with a
composite pulse, in the presence of errors, what a single pulse would have
performed in the absence of errors~\cite{Lev86}.  In the present case, the
goal is somewhat different: we construct a sequence of gates which take the
system into higher motional levels and back again to perform some logical
operation.  Nevertheless, the idea of stringing together several pulses to
perform a simple gate has proved fruitful.  It is our hope that further
progress can be made in the experimental implementation of quantum
information processing devices by considering how existing tools can be
applied to different physical implementations.

%%%%%%%%%%%%%%%%%%%%%%%%%%%%%%%%%%%%%%%%%%%%%%%%%%%%%%%%%%%%%%%%%%%%%%%%%%%

\end{multicols}


\begin{thebibliography}{10}
\bibitem{SigmaNote}
  Note that $\sigma^+$ corresponds to a transition from $|1\>$ to $|0\>$,
  which accounts for the convention that $\sigma^+ a$ is a blue transition.
\end{thebibliography}
\end{document}